\documentclass[prd,aps,floatfix,nofootinbib,preprint ,tightenlines,superscriptaddress]{revtex4}
\usepackage{dcolumn}
\usepackage{amsmath}
\usepackage{latexsym}
\usepackage{graphicx}
\usepackage{bm}

\def\svev#1{\left\langle #1\right\rangle}       

\newcommand{\bee}{\begin{equation}}
\newcommand{\ee}{\end{equation}}
\newcommand{\beea}{\begin{eqnarray}}
\newcommand{\eea}{\end{eqnarray}}

\begin{document}

\title{Remarks about weighted energy integrals over Minkowski spectral functions
from Euclidean lattice data}
\author{Thomas DeGrand}
\affiliation{
Department of Physics, University of Colorado,
        Boulder, CO 80309 USA}
\email{thomas.degrand@colorado.edu}

\date{\today}

\begin{abstract}
I make some simple observations about the calculation of weighted averages over energy of Minkowski space
spectral densities from weighted averages over time of Euclidean space  correlation functions,
measured in latice simulations. 
 The  correlator of two vector currents
is used as an example, where it appears that a determination of a weighted average of
 the spectral function near the rho pole  at the five per cent level
is possible from lattice simulations.
\end{abstract}

\maketitle

Finding connections between theoretical calculations done in Euclidean space
 and results of experiments done in Minkowski space is a longstanding problem in many areas of
physics and involves many approaches. Lattice studies of QCD and other related systems are no exception.
This short note describes a simple technique for extracting
 weighted averages over energy of Minkowski space
spectral densities from Euclidean space lattice correlation functions.
 Examples are motivated by calculations of the hadronic vacuum polarization contribution  to the muon anomalous magnetic moment
(with Ref.~\cite{Aoyama:2020ynm} as my primary reference)
though there are obvious  applications to many similar processes \cite{Hansen:2017mnd}.

When the Euclidean correlation $\Pi(Q)$ is related to the Minkowski space spectral function $\rho(\omega)$
 measured at energy $\omega$
 by a once-subtracted
dispersion relation, the connection
between a Euclidean space correlation function
defined at Euclidean time $t$, $G_E(t)$, and the  spectral function is
\bee
G_E(t) = \frac{1}{2\pi}\int_0^\infty d\omega [\omega^2\rho(\omega)] \exp(-\omega t).
\label{eq:vct}
\ee
Inverting Eq.~\ref{eq:vct} to predict $\rho(\omega)$ from $G_E(t)$ is a difficult problem.
However, it seems easy to compare a weighted average of $G_E(t)$ to a weighted average of $\rho(\omega)$,
\bee
\hat \rho(Q_0) \equiv \int_0^\infty R_E(Q_0,t)G(t)dt =  \int_0^\infty d\omega \rho(\omega)T(Q_0,\omega)
\label{eq:vsimpler}
\ee
with the connection
\bee
  T(\omega) = \frac{\omega^2}{2\pi}\int_0^\infty e^{-\omega t}R_E(Q_0,t) dt   .
\label{eq:vtvsint}
\ee
$Q_0$ is shorthand for  possible tunable parameter(s) in the weighting function.
The most prominent present-day example of such a connection is the calculation of the hadronic vacuum
polarization contribution  for the anomalous magnetic moment of the muon $a_\mu^{HVP}$.
The Euclidean weighting function  $R_E(Q_0,t)$ for $a_\mu$ is specified by a QED calculation.
But of course, one could imagine doing the weighting with any  function $R_E(Q_0,t)$. 

Each choice of $R_E(Q_0,t)$ amounts to its own (indirect) comparison
of theory ($G_E(t)$) with experiment ($\rho(\omega)$, processed into $\hat \rho(Q_0)$).
Families of related $R_E(Q_0,t)$'s can be combined into more extensive views of the spectral density.
I don't want to speculate on whether this could do a better job of probing the spectral function than the standard technique
of fitting $G_E(t)$ to a functional form with a set of parameters (masses and coupling constants) and then
continuing the fit function from Minkowski to Euclidean space. I just want to raise the possibility that 
analysis methods for  $a_\mu^{HVP}$ might have wider applications.

Some choices of $R_E$ are going to be more interesting than others, and a desirable
goal would be to find an $R_E((Q_0,t)$ whose $T(\omega)$ is peaked around some energy range.
To jump to the conclusion, the dominant feature of an $R_E(Q_0,t)$   which does that is a restriction to a range of $t$ values
$t_{min}<t<t_{max}$; the overall shape of  $R_E(Q_0,t)$ does not seem to be important for the
examples I display. And given what is published about the precision of $a_\mu^{HVP}$  lattice results,
it seems possible to make a lattice determination of a weighted average of $\rho(\omega)$ with enough accuracy to 
be phenomenologically interesting.
(I have in mind the few per cent tension in the $\pi\pi$ channel in the 0.6-0.9 GeV range described in Ref.~\cite{Aoyama:2020ynm},
between the KLOE experiment \cite{KLOE-2:2017fda} and other groups.)

  I think I am saying obvious things, but I haven't found a discussion of this approach
in the literature.

The idea described here is just a trivial variation on the ``coordinate space 
representation'' for $a_\mu^{HVP}$: there is
an implicit assumption that $R_E(t)$ is a smooth function of $t$, and replacing an integral over
continuous $t$ by a  sum over a set of discrete lattice points 
 is no different than replacing any continuous integral by a grid sum.
There is also a large literature proposing solutions to the ``inverse problem:'' given 
a  $G_E(t_i)$ defined at a set of discrete $t_i$ values,  various approaches have different criteria for
defining and constructing a  weighted $\hat\rho$.
Often, no smoothness assumptions go into the choice and in fact the $R_E(t_i)$'s found in the literature
are far from smooth. Recent references (a very incomplete set for this vast field)
 are Ref.~\cite{Hansen:2017mnd}, which uses the Backus - Gilbert method
\cite{Backus,NR}, related work by Ref.~\cite{Hansen:2019idp}, and Chebychev techniques by 
Refs.~\cite{Gambino:2020crt,Fukaya:2020wpp,Bailas:2020qmv}.

I will continue the note focussing on $a_\mu^{HVP}$.
There is a small literature associated with modifications to its $R_E$. Probably the most prominent one is
the ``intermediate window method'' of Ref.~\cite{RBC:2018dos}. It is a time - sliced version
of the $a_\mu^{HVP}$ weighting:
\bee
R_E(Q_0,t) = R_E^{a_\mu}(Q_0,t) [\Theta(t,t_{min},\Delta) - \Theta(t,t_{max}\Delta) ]
\label{eq:iw}
\ee
where $\Theta(t,t_0,\Delta)$ is a smoothed step function.
Another approach to weighting, called  ``finite energy sum rules,'' starts by writing a dispersion
relation for a reweighted  $\Pi_E(Q)$.
For a discussion, see Refs.~\cite{Davier:1998dz,Maltman:1998uzw}.

 To set conventions, we are interested in the correlator of two vector currents
\bee
\Pi(q)_{\mu\nu} = \int d^4 x e^{iqx} \svev{0 | J_\mu(x) J_\nu(0)|0}.
\ee
We remove the indices with a transverse projection,
\bee
\Pi_{\mu\nu} = [q_\mu q_\nu = g_{\mu\nu}q^2]  \Pi(q^2)
\ee
and then the spectral function $\rho(\omega)$ is proportional to the discontinuity of $\Pi$ across the 
real energy axis (setting $q_\mu=(\omega,\vec 0)$). It is also
 proportional to the R-ratio,
$R(\omega) = \sigma(e^+ e^- \rightarrow hadrons)/ \sigma(e^+ e^- \rightarrow \mu^+ \mu^-)$.
The standard lattice vector - vector  correlator contracts $\rho_{\mu\nu}$ against polarization
vectors $\epsilon^i_\mu \epsilon^j_\nu$ where
typically $\epsilon^i_\mu=(0,\vec \epsilon^i)$ is a unit vector. This means that
 in Eq.~\ref{eq:vct}  $\rho(\omega)=R(\omega)/(6\pi)$ and
\bee
G_E(t)= \sum_i \int d^3x \svev{J_i(\vec x, t)J_i(0,0)}
\ee
where $J_i(x,t)= e_q \bar \psi(x,t)\gamma_i \psi(x,t)$ for a quark of charge $e_q$ (in units of the electric charge).

\begin{figure}
\begin{center}
\includegraphics[width=0.8\textwidth,clip]{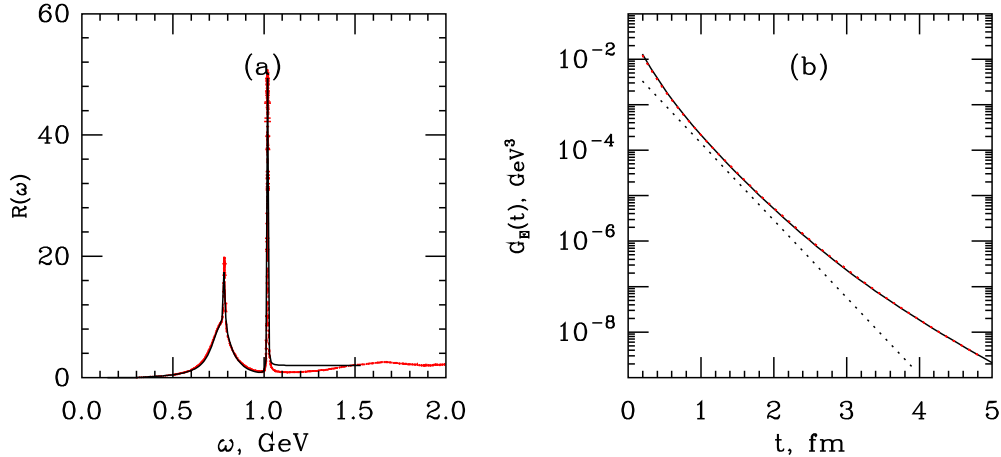}
\end{center}
\caption{ (a) $R(\omega)$ the R-ratio from the  phenomenological model for $\rho(\omega)$ from
 Ref.~\protect{\cite{Bernecker:2011gh}} (in black)  and from the table in the Review of Particle Properties
\protect{\cite{ParticleDataGroup:2020ssz}} (in red).
(b) $G_E(t)$ from the  phenomenological model for $\rho(\omega)$ from
 Ref.~\protect{\cite{Bernecker:2011gh}} (in black)  and from the table in the Review of Particle Properties
\protect{\cite{ParticleDataGroup:2020ssz}} (in red).
 The inversion uses Eq.~\protect{\ref{eq:vct}}. The dotted line is the contribution from a stable rho meson with $f_V=0.25$.
\label{fig:rdfcfm}}
\end{figure}

The  two relevant pictures are shown in Fig.~\ref{fig:rdfcfm}: the familiar plot of
 the R-ratio in  panel (a) and the 
 expected  $G_E(t)$ in panel (b),
 using Eq.~\ref{eq:vct} to do the inversion.
``Experiment'' in these pictures are  the  phenomenological model for $\rho(\omega)$ from
 Ref.~\cite{Bernecker:2011gh} (in black) and a compilation of $R(\omega)$ from 
a table in the Review of Particle Properties
\cite{ParticleDataGroup:2020ssz} (in red).
Of course, the  question to try to answer  is: Given a calculation $G_E(t)$, what can one say about $\rho(\omega)$?

This question is partially answered by the
 one theoretical line in panel (b) of Fig.~\ref{fig:rdfcfm}:
The straight line is the contribution of a stable rho meson at 770 MeV with a decay constant $f_V=0.25$:
\bee
G_E^V(t) = \frac{(\svev{q}m_V^2 f_V)^2}{2m_V} \exp(-m_V t) \label{eq:ct}.
\ee
The quantity $\svev{q}$ is the expectation value of the quarks' charges in the meson:
$[2/3-(-1/3)]/\sqrt{2}=1/\sqrt{2}$ for the rho, $1/18$ for the omega, $1/9$ for the phi, and so on.
$G_E(t)$ is flatter than $G_E^V(t)$  at very large $t$ due to the contribution of two-pion states with an invariant mass
smaller than the rho mass, and it is steeper than  $G_E^V(t)$ at small $t$ due to the phi meson and to the
flat high energy part of $R(\omega)$. Nowhere does $G_E^V(t)$ saturate $G_E(t)$.

I can rephrase the question to try to answer as:
Given a lattice calculation of $G_E(t)$, what can one say about $\rho(\omega)$? Then there are more constraints.
The large $\omega$ region, where $\omega> 1/a$ and $a$ is the lattice spacing, is contaminated by lattice artifacts,
and is inaccessible to a lattice calculation.
Unfortunately, so is the small $\omega$ or large $t$ region. There are two reasons for this. First,
 the lattice signal becomes noisy. This is a usual issue in 
lattice simulations \cite{Lepage:1989hd,Parisi:1983ae,Endres:2011jm}.
The data in Ref.~\cite{FermilabLattice:2019ugu} provide an example -- see their Fig.~2. The collaboration has data
at lattice spacings between 0.15 and 0.06 fm. Their data is only usable out to distances
 $t\sim 2.5$ fm. This precludes, at least for the present,
studies of $\rho(\omega)$ near threshold.
This situation is well known and documented in the $a_\mu^{HVP}$ literature \cite{Aoyama:2020ynm}.

A second reason that the small $\omega$ region is difficult is that it is dominated by two pion states.
Lattice simulations are done in a finite box (say, of size $L$) and particle momenta are quantized,
$\vec p_n =2\pi\vec n/L$ for integer valued $\vec n$,
 so that there is no two pion continuum in a lattice simulation, just a set of
exponentially falling contributions $\propto \exp(-2\sqrt{p_n^2+m_\pi^2} t)$. Presumably,
these contributions interpolate into the continuum result when the volume is taken sufficiently large,
Finite volume $\rho(\omega)$'s are sums of delta functions, but the smearing washes out this behavior.

Parenthetically, the vector correlator presents a somewhat special case compared to most lattice studies,
 where the lightest
state in (continuum) $\rho(\omega)$ is an isolated pole. Then, simply going to large $t$ gives a $G_E(t)$
which is dominated by properties of the pole. Standard lattice techniques (fits to exponentials)
are more efficient at producing high quality results than the proposal of weighting $G_E(t)$ given here.

So we are pushed back to the region of $\omega$ near the rho mass. The physical rho meson is broad.
 Is it possible to say anything about
$\rho(\omega)$ for $\omega$ near $m_\rho$? This seems to be a serious issue for $a_\mu^{HVP}$ determinations.

\begin{figure}
\begin{center}
\includegraphics[width=0.7\textwidth,clip]{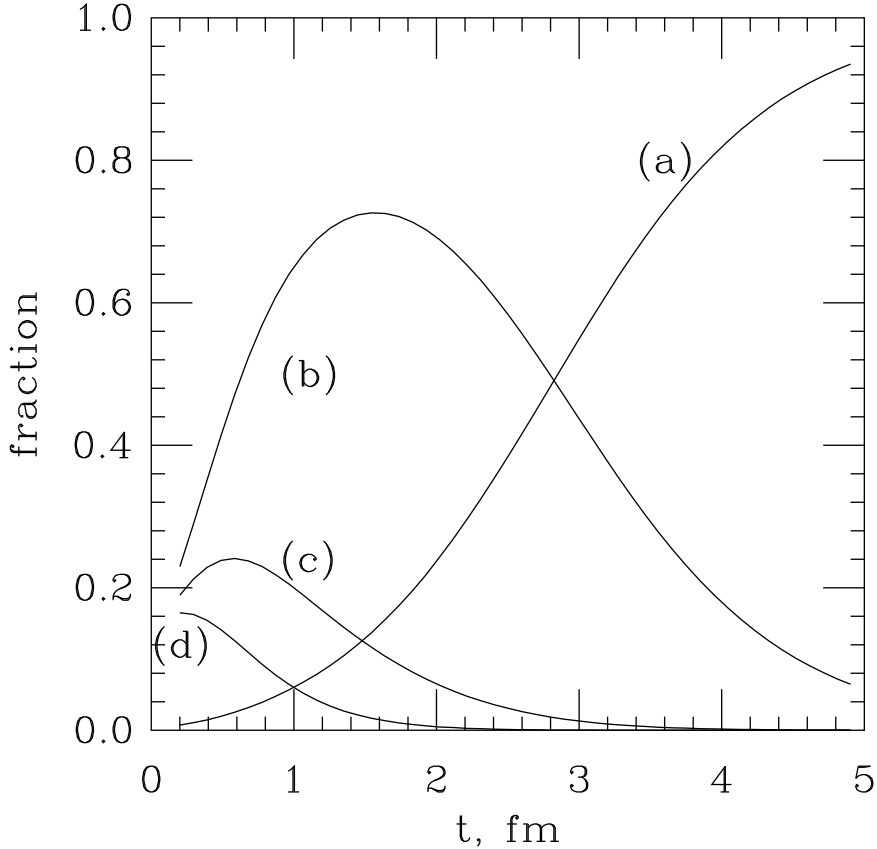}
\end{center}
\caption{ Fractional contributions to $G_E(t)$ where $\rho(\omega)$ is taken 
 from the  phenomenological model  of
 Ref.~\protect{\cite{Bernecker:2011gh}}. The curves  label
(a) $2m_\pi < \omega < 4m_\pi$;
(b) $4m_\pi < \omega < 6m_\pi$;
(c) $6m_\pi < \omega < 8m_\pi$;
(d) $8m_\pi < \omega < 10m_\pi$.
\label{fig:frac}}
\end{figure}

Dividing up the contributions to $G_E(t)$ from different energy intervals shows the way to go.
See Fig.~\ref{fig:frac}, which shows the fractional contributions to $G_E(t)$ from different $\omega$ regions.
Here $\rho(\omega)$ is taken 
 from the  phenomenological model of Ref.~\cite{Bernecker:2011gh}.
What is noticable is that there is a fairly wide region 
at intermediate $t$ where the region around the rho mass contributes heavily. Of course, this can be seen by eye
in Fig.~\ref{fig:rdfcfm}. (This is basically just the phenomenon of vector meson dominance.)

As an application of this remark, suppose that the experimental $\rho(\omega)$ is
not precisely known over some energy range, that two 
experiments differ by a fraction $\delta \rho(\omega)/\rho(\omega)$. Assuming that the
difference is confined to some small region of $\omega$, there will be
a change in $G_E(t)$  (constructing it from Eq.~\ref{eq:vct} with each experimental $\rho(\omega)$)
of $\delta G_E(t)/G_E(t)\sim f \delta \rho/\rho$ where $f$ is the fractional
contribution of the $\omega$ region of $\rho(\omega)$ to $G_E(t)$.

\begin{figure}
\begin{center}
\includegraphics[width=0.7\textwidth,clip]{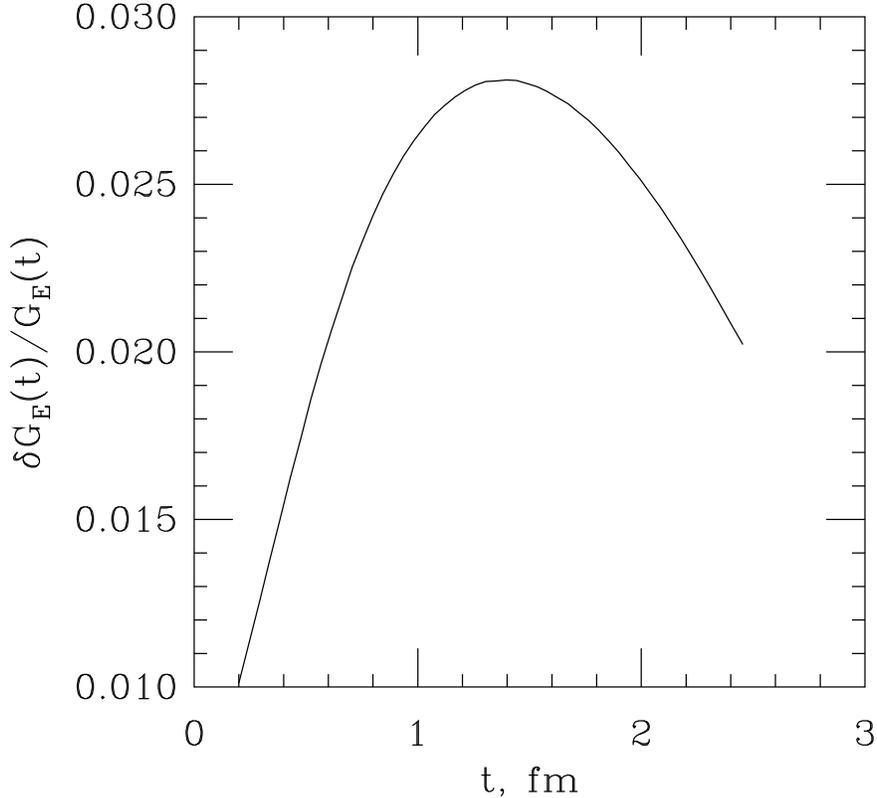}
\end{center}
\caption{ The fractional change in $G_E(t)$ from a five per cent variation in 
the 
 phenomenological model for $\rho(\omega)$ of
 Ref.~\protect{\cite{Bernecker:2011gh}} over the range 0.6-0.9 GeV, as described in the text.
\label{fig:fracct}}
\end{figure}

 A simple example comes from modifying the model for $\rho(\omega)$ from
 Ref.~\cite{Bernecker:2011gh} over a range $\omega_{min} < \omega < \omega_{max}$,
by multiplication by a weighting factor
\bee
w(\omega)=1 + a \sin\pi \left( \frac{\omega-\omega_{min}}{\omega_{max}-\omega_{min}}\right).
\label{eq:modelwt}
\ee
The fractional change in $G_E(t)$ is shown in Fig.~\ref{fig:fracct}
for the choice $\omega_{min}=0.6$ GeV, $\omega_{max}=0.9$ GeV, $a=0.05$.

Notice the qualifier ``assuming  that the
difference is confined to some small region of $\omega$.'' $G_E(t)$ at any $t$ value is built of contributions 
from all $\omega$, and a measurement of $G_E(t)$ at any $t$ or for any range of $t$ values does not
make an absolute prediction about $\rho(\omega)$ at any particular $\omega$ value.
However, lattice results could still be useful to distinguish between the different experimental $\rho(\omega)$'s.

Fig.~\ref{fig:fracct} shows that a five per cent variation in $\rho(\omega)$ translates 
into a 2.7 per cent variation in $G_E(t)$
over a fairly wide range of $t$. This is really the end of the story I can tell -- I cannot write about uncertainties
in either $G_E(t)$ or $\rho(\omega)$. Lattice data for $G_E(t)$ is typically highly correlated
and it is almost impossible to estimate correlation uncertainties in a lattice data set without
access to it. Similarly, the experimental data sets which give $\rho(\omega)$ are highly correlated.
 But, 2.7 per cent seems to be an easy target,
 given that contemporary lattice measurements of $a_\mu^{HVP}$ are well under a per cent. It seems likely that
lattice calculations could determine $\rho(\omega)$ over the range 0.6-0.9 GeV at the five per cent level.

To do this in practice, we need a weighting function. There seem to be many possible choices. But
Figs.~\ref{fig:frac} and \ref{fig:fracct} indicate that
all that is significant for $R_E(t)$  is that
it has a cutoff at small and large $t$, that it is nonzero for $t_{min}<t<t_{max}$.
 Two choices of $R_E(t)$ illustrate that claim.

\begin{figure}
\begin{center}
\includegraphics[width=0.8\textwidth,clip]{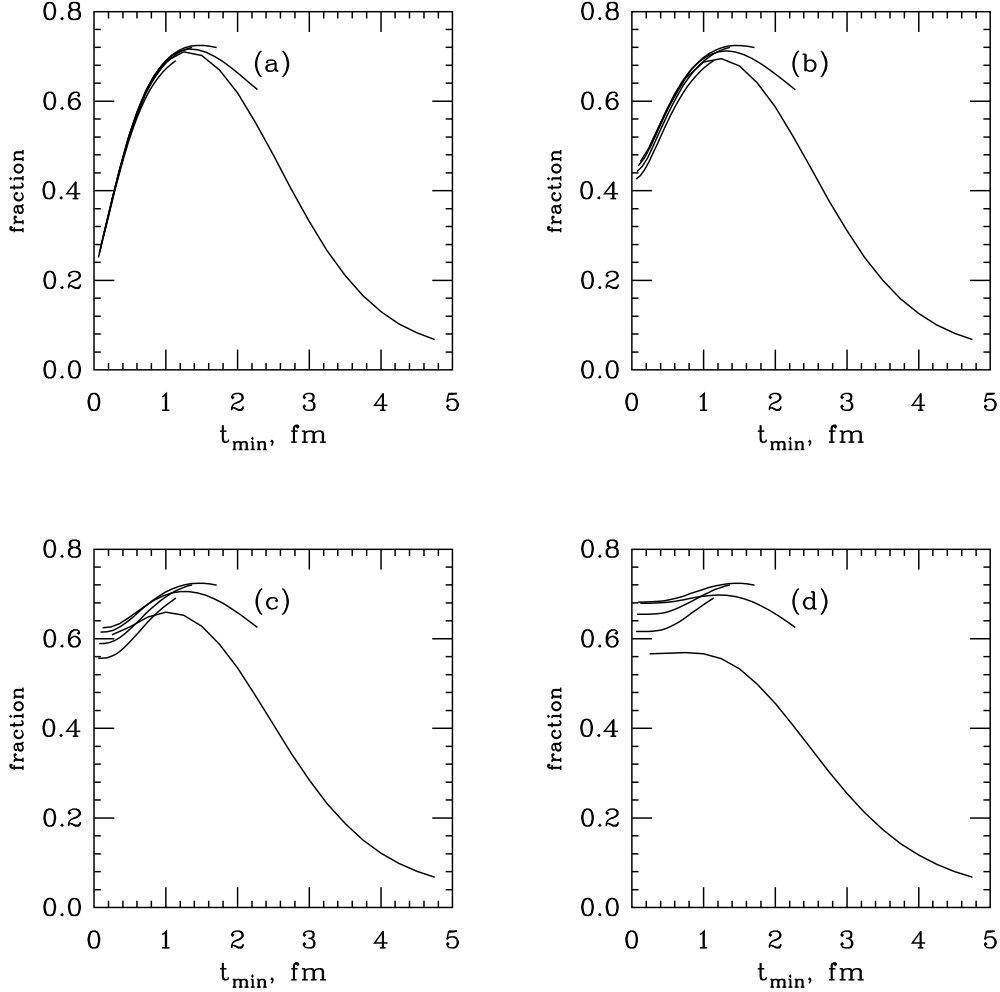}
\end{center}
\caption{ The contribution of $4m_\pi < \omega < 6m_\pi$ to the integral of Eq.~\protect{\ref{eq:vsimpler}}
for a power law $R_E(t)=(t/t_0)^n/n!$ with $t_0=0.15$ fm, for a range $t_{min}<t<t_{max}$ plotted versus $t_{min}$
for $t_{max}=1.2$, 1.44, 1.8, 2.4 and 5 fm.
(a) $n=0$;
(b) $n=2$;
(c) $n=4$;
(d) $n=6$.
\label{fig:fracpow}}
\end{figure}

First consider a family of power laws,  $R_E(t)=(t/t_0)^n/n!$ for a range $t_{min}<t<t_{max}$. $t_0$ 
and the $n!$ factor are just  rescalings, useful for plots across $n$ or for comparing 
weighted lattice data at different lattice spacings. (This
 is a hard cutoff; Fig.~\ref{fig:fracct} indicates that a soft
cutoff would perform similarly.)
Fig.~\ref{fig:fracpow} shows the contribution of $4m_\pi < \omega < 6m_\pi$ to the integral of Eq.~\ref{eq:vsimpler}.
 Each panel is for a particular $n$ value and shows a set of curves: each curve  is the fraction of the integral from
$t_{min}$ to $t_{max}$ from this $\omega$ range, varying $t_{min}$ at fixed $t_{max}$. A range of $t$ in the range 1-2 fm
gives an integral where the contribution of
 $4m_\pi < \omega < 6m_\pi$ region of $\rho(\omega)$ approaches 70 per cent, essentially independent of $n$.
The curves extending out to $t_{max}=5$ fm show the obvious result that the contribution of the rho region
to the integral becomes very small when taking $t_{min}>2$ fm.

\begin{figure}
\begin{center}
\includegraphics[width=0.8\textwidth,clip]{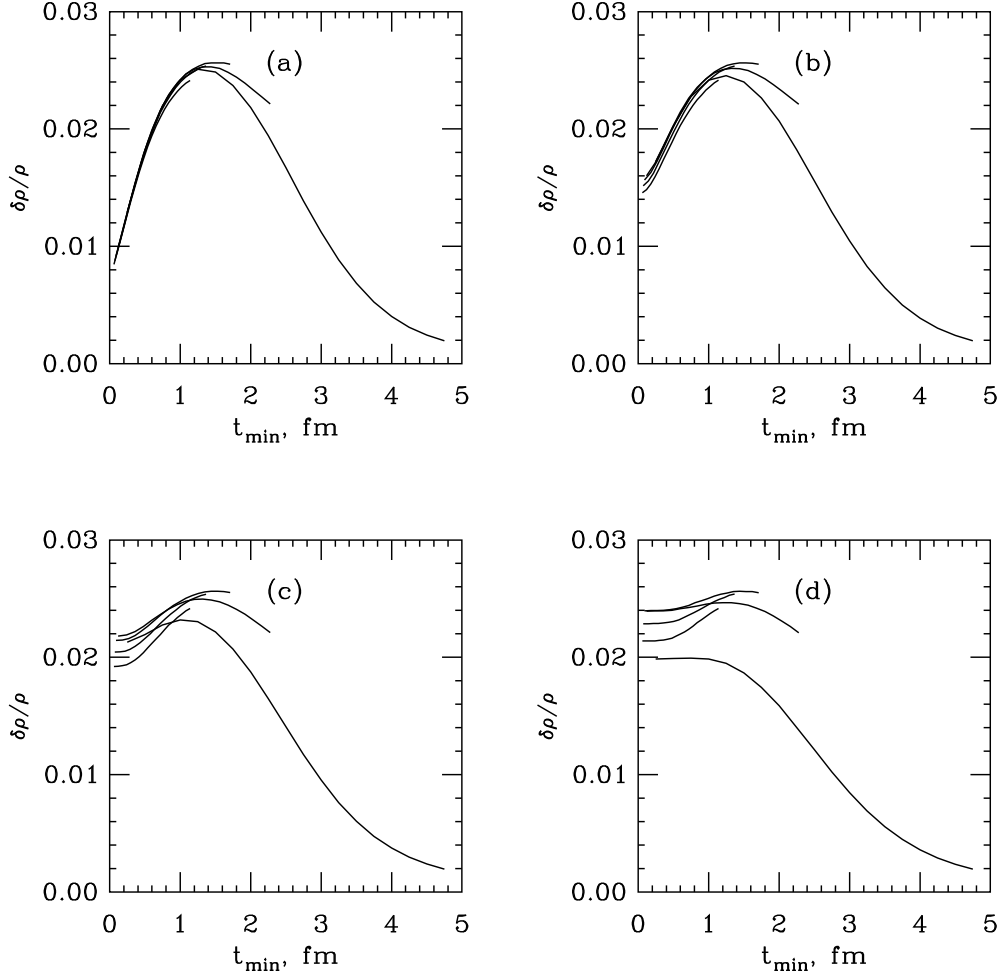}
\end{center}
\caption{ Fractional change in the integral $\hat \rho$ of Eq.~\protect{\ref{eq:vsimpler}}
 under a five per cent variation in $\rho(\omega)$
for $4m_\pi < \omega < 6m_\pi$ parameterized as in Eq.~\protect{\ref{eq:modelwt}},
for a power law $R_E(t)=(t/t_0)^n/n!$ with $t_0=0.15$ fm,   for a range $t_{min}<t<t_{max}$ plotted versus $t_{min}$
for $t_{max}=1.2$, 1.44, 1.8, 2.4 and 5 fm.
(a) $n=0$;
(b) $n=2$;
(c) $n=4$;
(d) $n=6$.
\label{fig:fracdr}}
\end{figure}

Fig. ~\ref{fig:fracdr} shows the fractional change in the integral $\hat \rho$ from the model weighting factor
of Eq.~\ref{eq:modelwt}. This is the analog of Fig.~\ref{fig:fracct} and the result is the same -- the sensitivity
to variation in $\rho(\omega)$ depends most on the range of $t$ spanned by $R_E(t)$.

\begin{figure}
\begin{center}
\includegraphics[width=0.8\textwidth,clip]{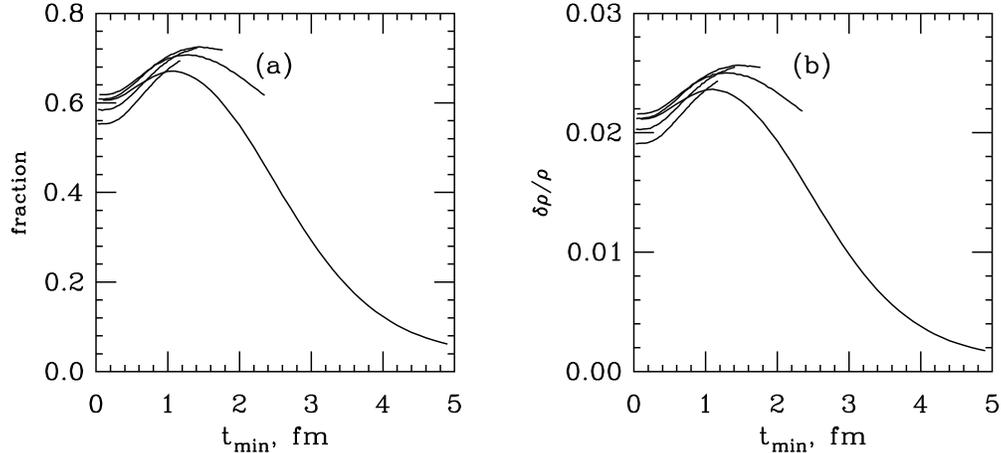}
\end{center}
\caption{ The analog of Figs.~\protect{\ref{fig:fracpow}} and \protect{\ref{fig:fracdr}}
 for the smearing kernel for $a_\mu^{HVP}$.
Panel (a) is the
 contribution of $4m_\pi < \omega < 6m_pi$ to the integral of Eq.~\protect{\ref{eq:vsimpler}}.
Panel (b) is the fractional  change in the integral $\hat \rho$ of Eq.~\protect{\ref{eq:vsimpler}}
 under a five per cent variation in $\rho(\omega)$ 
for $4m_\pi < \omega < 6m_\pi$ parameterized as in Eq.~\protect{\ref{eq:modelwt}}. Again, the curves are
 $t_{min}<t<t_{max}$ plotted versus $t_{min}$ for
$t_{max}=1.2$, 1.44, 1.8, 2.4 and 5 fm.
\label{fig:fraciw}}
\end{figure}

Fig.~\ref{fig:fraciw} shows similar results for the smearing kernel used for $a_\mu^{HVP}$ in its intermediate window
guise. The figures are nearly identical to the ones for power law weighting.
The conclusion seems to be that a lattice calculation of $G_E(t)$ with an accuracy of 2-3 per cent
(in the continuum limit, of course) over the range of 1-2 fm 
can distinguish a five per cent  variation in $\rho(\omega)$ in the rho region.

At this point I should stop and hope for an analysis by one of the lattice groups using its own  data sets.
The idea I have presented is trivial, but it also seems simple to implement.
I think that $\rho(\omega)$ (and related quantities) are  interesting in and of themselves, and that
trying to extract features of $\rho(\omega)$ which have nothing to do with $a_\mu^{HVP}$  from $G_E(t)$ could
be a useful project. And, of course, identical weighting techniques can  connect other
inclusive processes with Euclidean correlators.

\begin{acknowledgments}
Questions at the end of a seminar
by Finn Stokes at the MIT Virtual Lattice Colloquium Series motivated this note.
I would also like to thank Maarten Golterman for correspondence and Chris Polly for encouragement.
This material is based upon work supported by the U.S. Department of Energy, Office of Science, Office of
High Energy Physics under Award Number DE-SC-0010005.
\end{acknowledgments}

\end{document}